# Pure spin photocurrent in non-centrosymmetric crystals: bulk spin photovoltaic effect

Haowei Xu [1], Hua Wang [1], Jian Zhou [1], and Ju Li [1,2 †]

[1] Department of Nuclear Science and Engineering, Massachusetts Institute of Technology, Cambridge, Massachusetts 02139, USA

[2] Department of Materials Science and Engineering, Massachusetts Institute of Technology, Cambridge, Massachusetts 02139, USA

## Abstract

Spin current generators are critical components for spintronics-based information processing. In this work, we theoretically and computationally investigate the bulk spin photovoltaic (BSPV) effect for creating DC spin current under light illumination. The only requirement for BPSV is inversion symmetry breaking, thus it applies to a broad range of materials and can be readily integrated with existing semiconductor technologies. The BSPV effect is a cousin of the bulk photovoltaic (BPV) effect, whereby a DC charge current is generated under light. Thanks to the different selection rules on spin and charge currents, a pure spin current can be realized if the system possesses mirror symmetry or inversion-mirror symmetry. The mechanism of BPSV and the role of the electronic relaxation time $\tau$ are also elucidated. We apply our theory to several distinct material systems, including transition metal dichalcogenides, anti-ferromagnetic $MnBi_2Te_4$, and the surface of topological crystalline insulator cubic SnTe.

---

[†] correspondence to: liju@mit.edu. ORCID of Ju Li: https://orcid.org/0000-0002-7841-8058

## Introduction

Present-day electronics, which utilize the charge degree of freedom of electrons, have revolutionized human civilization. Besides charge, spin is another intrinsic freedom of electrons that can be exploited for information processing. Indeed, spintronics[1,2] is promising for next-generation energy-efficient devices and other novel applications such as quantum computing[3,4] and neuromorphic computation[5]. One of the core challenges[6] of spintronics is the generation of a spin current, and particularly, a *pure* spin current without an accompanying charge current. Until now, there have been a few approaches, such as the interconversion between charge and spin currents by (inverse) spin galvanic effect[7,8] or (inverse) spin Hall effect[9–12], and the interconversion between thermal and spin currents by spin Seebeck effect[13,14] or spin Nernst effect[15,16], etc. These approaches require electrode contact and patterning, and the response time is usually on the order of nanoseconds or longer. In contrast, optical approaches are non-contact, non-invasive, and can enable ultrafast response time on the order of picoseconds and below.

Several optical approaches for generating spin currents have been proposed, however, these approaches typically require special ingredients, such as the breaking of time-reversal symmetry $\mathcal{T}$ by introducing magnetic elements or circularly polarized light (CPL), and/or special device structures. For example, CPL can selectively couple with spin-up and down states in quantum wells[17], or spin-valley locked systems[18], and the imbalanced population of spin-up and down states could lead to a spin photocurrent. In magnetic element material systems, it has also been proposed that a linearly polarized light (LPL) can generate a spin current with the shift-current mechanism[19–22]. Alternatively, a spin current can be generated with a mechanism reminiscent of the p-n junction in solar cells[23–25], quantum interference[26,27], or the nonlinear Drude current[28]. Although progress has been made, the generation of spin currents under light is still underexplored. In particular, it is highly desirable to introduce new mechanisms applicable to a broader family of materials.

In this work, we propose mechanisms to generate DC spin current based on the nonlinear optical (NLO) theory, which is a cousin of the bulk photovoltaic (BPV) effect[29,30], whereby DC charge currents can be generated in a uniform crystal under light illumination. The BPV, together with other NLO effects, are under intensive research recently, but thus far the attention is mainly on the charge current, while the spin current has long been neglected. Certainly, when the charge flows under light, the spin associated with the carriers are moving as well, which is a spin current. In some situations, the charge current vanishes due to symmetry, but this does not indicate that the carriers are frozen in materials. Indeed, the carriers generally still move under above-band-gap illumination, which leads to a nonzero pure spin current. A generic picture here is that electrons with opposite (or at least different) spin polarizations travel in the opposite directions so that the net charge current is zero while the net spin current is nonzero (Figure 1). We name this effect the Bulk Spin Photovoltaic (BSPV) effect. Here the "voltaic" is defined as $V_{\uparrow\downarrow} \equiv (\mu_\uparrow - \mu_\downarrow)/(-e)$, which is the difference in the chemical potential of spin-up ($\mu_\uparrow$) and spin-down ($\mu_\downarrow$) electrons. This should be compared with the BPV voltage, which may be defined as $U \equiv (\mu_\uparrow + \mu_\downarrow)/(-2e)$. Similar to the BPV voltage $U$, $V_{\uparrow\downarrow}$ will not be limited by the bandgap of the material, and the currents will not be limited by the Shockley–Queisser limit.

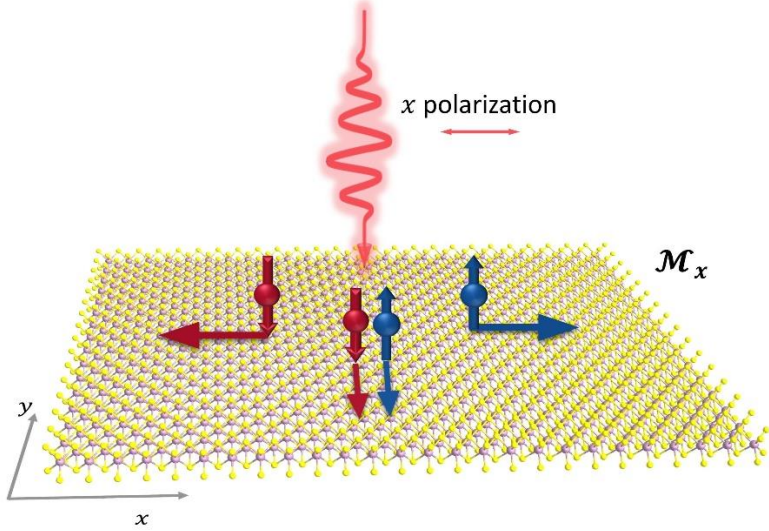

**Figure 1**. A schematic illustration of pure spin and charge current. The light polarizes in the $x$ direction, while the system has mirror symmetry $\mathcal{M}_x$. In the $x$ direction, spin up and down states travel in opposite directions, so that the net charge current is vanishing, whereas the net spin current goes to the $+x$ direction. In the $y$ direction, spin up and down electrons travel in the same direction, leading to non-vanishing charge current but vanishing spin current.

In the following, we first introduce a unified theory on NLO spin (BSPV) and charge (BPV) currents generation. Then, combining theoretical analysis and ab initio calculations, we elucidate some prominent features of the BSPV. Notably, the only requirements for BSPV are (a) above-direct-band-gap light illumination, and (b) the breaking of inversion symmetry $\mathcal{P}$, regardless of $\mathcal{T}$. There is no need for any special ingredients such as magnetic materials, special device structures (quantum wells, junctions, etc.), the interference between two pulses, or specific light wavelength or polarization. Hence, BSPV has great convenience in practice and can be readily integrated with existing semiconductor technologies[31,32]. These advantages, together with the flexibilities of optical approaches (dynamic spatial addressability, tunable intensity, wavelength, polarization, etc.), provide a large playground to be explored. These results are useful not only for generating spin currents but also for material characterization and sensing. Many applications that are not envisaged before may become possible. In addition, we also clarify the lattice symmetry requirements for the generation of pure spin current, and the mechanisms (shift- and/or injection-like) for spin current generation under different symmetry conditions and light polarizations.

## Results

**General Theory and Symmetry Analysis.** The NLO charge or spin current under light with frequency $\omega$ can be expressed as

$$J^{a,s^i} = \sum_{\Omega=\pm\omega} \sigma^{a,s^i}_{bc}(0;\Omega,-\Omega) E^b(\Omega) E^c(-\Omega) \tag{1}$$

Here $E(\omega)$ is the Fourier component of the electric field at angular frequency $\omega$. $\sigma^{a,s^i}_{bc}$ is the NLO conductivity, with $a,b,c$ as Cartesian indices. $a$ indicates the direction of the current, while $b$ and $c$ are the polarization of the optical electric field. $s^i$ with $i=x,y,z$ is the spin polarization, while $s^0$ represents charge current. The spin and charge are in the unit of angular momentum $\frac{\hbar}{2}$ and electron charge $e$, respectively. To directly compare the values of the charge and spin current conductivity, we divide the spin current conductivity by a factor of $\frac{\hbar}{2e}$ [33]. Equation (1) suggests that the $+\omega$ and $-\omega$ components of the electric field are combined, and a static current is generated. We derived the formula for $\sigma^{a,s^i}_{bc}$ from quadratic response theory[30,34] (see Supplementary Materials, SM). Within the independent particle approximation, the conductivity can be expressed as

$$\sigma^{a,s^i}_{bc}(0;\omega,-\omega) = -\frac{e^2}{\hbar^2\omega^2}\int\frac{d\mathbf{k}}{(2\pi)^3}\sum_{mnl}\frac{f_{lm}v^b_{lm}}{\omega_{ml}-\omega+i/\tau}\left(\frac{j^{a,s^i}_{mn}v^c_{nl}}{\omega_{mn}+i/\tau} - \frac{v^c_{mn}j^{a,s^i}_{nl}}{\omega_{nl}+i/\tau}\right) \tag{2}$$

Here the explicit $k$-dependence of the quantities are omitted. $f_{lm} = f_l - f_m$ and $\omega_{lm} = \omega_l - \omega_m$ are the difference of occupation number and band energy between bands $l$ and $m$. $v_{nl} \equiv \langle n|\hat{v}|l\rangle$ is the velocity matrix element, while $\tau$ is the carrier lifetime, and is set to be 0.2 ps uniformly in this paper. The symmetric real and asymmetric imaginary part of $\sigma_{bc}^{a,s^i}$ correspond to the conductivity under LPL and CPL, respectively. Note that Equation (2) uses the velocity gauge, while the well-known shift and injection charge current formulae[35] use the length gauge. These two gauges are equivalent[36,37] (SM). An advantage of the velocity gauge is that the equations are relatively short and neat, and are easily generalizable to other responses under light, such as valley currents, static magnetization, etc.

The physical mechanism of BSPV can be better understood when compared with BPV. In Equation (2), $j^{a,s^i}$ with $i \neq 0$ is the spin current operator, defined as[38] $j^{a,s^i} = \frac{1}{2}(v^a s^i + s^i v^a)$. Here $s^i = \frac{\hbar}{2}\sigma^i$ is the spin operator with $\boldsymbol{\sigma}$ as the Pauli matrices. Note that there are lots of debates on the definition of spin current[39–41], see SM for detailed discussions. If we define $s^0 = e$, then $j^{a,s^0}$ would indicate the charge current in BPV. The unified formula for spin and charge currents indicates that the DC spin current has a similar physical picture as the BPV, except that spin is a pseudovector, thus it has different symmetries and selection rules from the charge, which is a scalar. When electron moves, its charge and spin would move simultaneously, leading to the charge and spin current, respectively. However, unlike charge, which is always $-|e|$ for an electron, spin does not necessarily have a specified value. A free electron can have equal probability to have $s_z = \frac{1}{2}$ or $-\frac{1}{2}$. When free electrons move to the right, the spin-$z$ current associated would have an equal probability to be in the right (when $s_z = \frac{1}{2}$) or the left (when $s_z = -\frac{1}{2}$) direction, and the net spin current is thus zero.[42] Therefore, BSPV requires that the electrons have specified spin polarizations (i.e., a spin texture), which can be created by either spin-orbit coupling (SOC), or intrinsic magnetic ordering. Different from the formalism used in Ref. [21], Equation (2) does not require the spin to be a good quantum number or treat spin up and down states separately, so it can deal with arbitrary spin polarization under SOC. Later we will show that treating SOC in such a rigorous way is of importance.

**Table 1** The behavior of physical quantities under symmetry operations. Here $\tilde{\cdot}$ indicates $\cdot$ obtained on the $\mathcal{PT}$ partner state, which is degenerate in energy with the original state. $[d; abc]$ is $-1$ and $+1$ if there are odd and even numbers of $d$ within $a, b$ and $c$. For example, $[x; xxx] = -1$ while $[x; xxy] = 1$. $\boldsymbol{k}' = \mathcal{M}^d \boldsymbol{k}$ is the mirror image of $\boldsymbol{k}$ (only the $d$-th component of $\boldsymbol{k}$ is flipped).

|  | $v_{mn}^a(\boldsymbol{k})$ | $s^i(\boldsymbol{k})$ $(i \neq 0)$ | $N_{mnl}^{0abc}(\boldsymbol{k})$ | $N^{iabc}(\boldsymbol{k})$ $(i \neq 0)$ |
|---|---|---|---|---|
| $\mathcal{P}$ | $-v_{mn}^a(-\boldsymbol{k})$ | $s_{mn}^i(-\boldsymbol{k})$ | $-N_{mnl}^{0abc}(-\boldsymbol{k})$ | $-N_{mnl}^{iabc}(-\boldsymbol{k})$ |
| $\mathcal{T}$ | $-v_{mn}^{a*}(-\boldsymbol{k})$ | $-s_{mn}^{i*}(-\boldsymbol{k})$ | $-N_{mnl}^{0abc*}(-\boldsymbol{k})$ | $N_{mnl}^{iabc*}(-\boldsymbol{k})$ |
| $\mathcal{PT}$ | $\tilde{v}_{mn}^{a*}(\boldsymbol{k})$ | $-\tilde{s}^{i*}_{mn}(\boldsymbol{k})$ | $\tilde{N}_{mnl}^{0abc*}(\boldsymbol{k})$ | $-\tilde{N}_{mnl}^{iabc*}(\boldsymbol{k})$ |
| $\mathcal{M}^d$ | $(-1)^{\delta_{da}} v_{mn}^a(\boldsymbol{k}')$ | $-(-1)^{\delta_{di}} s_{mn}^i(\boldsymbol{k}')$ | $[d;abc] \times N_{mnl}^{0abc}(\boldsymbol{k}')$ | $-(-1)^{\delta_{di}} [d;abc] \times N_{mnl}^{iabc}(\boldsymbol{k}')$ |
| $\mathcal{PM}^d$ | $-(-1)^{\delta_{ab}} v_{mn}^a(-\boldsymbol{k}')$ | $-(-1)^{\delta_{di}} s_{mn}^i(-\boldsymbol{k}')$ | $-[d;abc] \times N_{mnl}^{0abc}(-\boldsymbol{k}')$ | $(-1)^{\delta_{di}} [d;abc] \times N_{mnl}^{iabc}(-\boldsymbol{k}')$ |

Next, we consider symmetry constraints on the conductivity tensor. First, the numerators in Equation (2) are composed of terms with the format $N_{mnl}^{iabc} = j_{mn}^{a,s^i} v_{nl}^b v_{lm}^c$ ($i \neq 0$) for spin current and $N_{mnl}^{0abc} = v_{mn}^a v_{nl}^b v_{lm}^c$ ($i = 0$) for charge current. Under spatial inversion operation $\mathcal{P}$, one has $\mathcal{P} v_{mn}^a(\boldsymbol{k}) = -v_{mn}^a(-\boldsymbol{k})$, $\mathcal{P} s_{mn}^i(\boldsymbol{k}) = s_{mn}^i(-\boldsymbol{k})$, and $\mathcal{P} j_{mn}^{a,s^i}(\boldsymbol{k}) = -j_{mn}^{a,s^i}(\boldsymbol{k})$. Thus, $\mathcal{P} N_{mnl}^{iabc}(\boldsymbol{k}) = -N_{mnl}^{iabc}(-\boldsymbol{k})$ for both $i \neq 0$ and $i = 0$. On the other hand, the denominators in Equation (2) are invariant under $\mathcal{P}$, thus all components (including charge and spin) of $\sigma_{bc}^{a,s^i}$ should vanish after a summation over $\pm \boldsymbol{k}$ in $\mathcal{P}$-conserved systems. Therefore, the inversion symmetry $\mathcal{P}$ has to be broken for both BPV and BSPV. Regarding time-reversal operation $\mathcal{T}$, one has $\mathcal{T} v_{mn}^a(\boldsymbol{k}) = -v_{mn}^{a*}(-\boldsymbol{k})$ and $\mathcal{T} s_{mn}^i(\boldsymbol{k}) = -s_{mn}^{i*}(-\boldsymbol{k})$ ($i \neq 0$. Here $\cdot^*$ indicates complex conjugate of $\cdot$). For charge current, one has $\mathcal{T} N_{mnl}^{0abc}(\boldsymbol{k}) = -N_{mnl}^{0abc*}(-\boldsymbol{k})$. Thus, the real and imaginary part of $N_{mnl}^{0abc}$ are odd and even under $\mathcal{T}$, respectively. The imaginary part of $N^{0abc}(\boldsymbol{k})$ contributes to the total charge conductivity after the summation over $\pm \boldsymbol{k}$ in a $\mathcal{T}$-conserved system. Similarly, for spin-$i$ current ($i \neq 0$), one has $\mathcal{T} N_{mnl}^{iabc}(\boldsymbol{k}) = N_{mnl}^{iabc*}(-\boldsymbol{k})$, thus it is the real part of $N^{iabc}(\boldsymbol{k})$ that contributes to the total spin conductivity. For both charge and spin current, $\mathcal{T}$ does not need to be broken. Generally speaking, spin and charge currents should be generated simultaneously in the absence of $\mathcal{P}$. However, as we will show in detail later, a pure spin current can be realized if the system possesses mirror symmetry $\mathcal{M}^d$, inversion-mirror symmetry $\mathcal{PM}^d$ or inversion-spin rotation symmetry $\mathcal{PS}$. The behavior of relevant physical quantities under different symmetry operations is summarized in Table 1.

The carrier lifetime $\tau$ plays a rather important role. Here we use the charge current as the example, a similar analysis applies to the spin current. The DC photocurrent is basically $j^a = \sigma_{bc}^a E^b E^c$. If the system is non-magnetic, and we use LPL, then it seems that $\mathcal{T}$ should be preserved. In this case, seemingly $\sigma_{bc}^a$ should be zero, because the $j^a$ is odd under $\mathcal{T}$, while $E^b E^c$ is even. However, in practice the nonlinear photocurrent does exist, which is the BPV (shift current). In fact, here $\mathcal{T}$ is effectively broken by dissipation in the thermodynamic second-law sense, characterized by $\tau$. This is related to the well-known paradox regarding microscopic reversibility: if particles in a movie satisfy Newton's equations of motion, then its rewinding version ($t \to -t$) would also, thus the apparent time-reversal symmetry in the equation of motion. However, if one watches the two movies ($+t$ and $-t$) for long enough time, then the "real" movie is the one with an overall "neater arrangement" of particles at the beginning of play, due to asymmetry in the initial condition. In other words, entropy creation indicates the "arrow of time" and distinguishes between $t$ and $-t$. Therefore, it has been rationalized that the electronic relaxation time $\tau$ is indispensable for the shift current, although the shift-current conductivity $\sigma_{bc}^a$ is independent of $\tau$.[35]

Dissipation occurs by the scattering of electrons and holes with phonons, etc., which lead to electron-hole recombination. The scattering time $\tau$ is usually on the order of (sub)-picoseconds. In some cases, the spin relaxation time is short, then it can be a source of dissipation as well. Also, in the presence of scattering potentials (from e.g., impurities), there could be skew scattering[43,44] and side jump[45,46], which lead to extrinsic spin/charge currents, as compared with the intrinsic currents studied in this work, that originates from the intrinsic band structure of the perfect crystal. Here we adopt the constant relaxation time approximation and use a constant $\tau = 0.2$ ps for all modes (band index $n$ and wavevector $k$). In reality $\tau$ should be mode dependent (see SM for more discussions) of course. This however does not affect the qualitative features of the theory.

To illustrate the theory, we investigate three distinct material systems: (1) monolayer transition metal dichalcogenides (TMD) which are $\mathcal{P}$-broken but $\mathcal{T}$-preserved; (2) antiferromagnetic bilayers MnBi$_2$Te$_4$ which is $\mathcal{P}$- and $\mathcal{T}$-broken but $\mathcal{PT}$-preserving; (3) the {001} surface of cubic SnTe which is $\mathcal{P}$-broken but has double mirror symmetry $\mathcal{M}_x$ and $\mathcal{M}_y$. The results suggest that BSPV is generic and robust in these distinct systems. We only show the NLO charge and spin current under LPL, while the responses under CPL can be found in the SM.

**Monolayer Transition Metal Dichalcogenide.** 2H-phase TMDs are well-studied 2D materials that possess many exotic electronic and optical properties. We take monolayer 2H MoS$_2$ as an example. The atomic structure of monolayer 2H MoS$_2$ (space group $P\bar{6}m2$) is shown in the inset of Figure 2(e), which lacks $\mathcal{P}$, but is invariant under $\mathcal{M}^x$ and $\mathcal{M}^z$. Monolayer TMDs exhibit Zeeman-type (out-of-plane) spin splitting due to the in-plane anisotropy. This could be understood with the effective magnetic field from SOC, expressed as $\boldsymbol{B}_{\text{eff}} = \frac{1}{2m_e c^2} \boldsymbol{p} \times \boldsymbol{\nabla} V$, where $m_e$ is the electron mass and $c$ is the speed of light. In monolayer TMDs, the momentum $\boldsymbol{p}$ is in the in-plane ($x$-$y$) direction, while $\boldsymbol{\nabla} V$ is also largely in the $x$-$y$ plane, due to the mirror plane $\mathcal{M}^z$. As a result, $\boldsymbol{B}_{\text{eff}}$ is mainly along the out-of-plane direction, leading to the Zeeman type spin splitting. These arguments are verified by the spin texture $s^i_{mm}(\boldsymbol{k}) = \langle m\boldsymbol{k}|\sigma_i|m\boldsymbol{k}\rangle$ from ab initio calculations. Figures 2(a) and 2(b) show $s^z_{mm}(\boldsymbol{k})$ for the highest valence band and the lowest conduction band of MoS$_2$, respectively. One can see that $s^z_{mm}(\boldsymbol{k}) \cong \pm 1$ for nearly all $\boldsymbol{k}$-points. Also, $s^z_{mm}(\boldsymbol{k})$ is opposite near the K and K' valleys, which is the spin-valley locking[47,48].

Here we need to examine constraints on NLO spin or charge current from mirror symmetry $\mathcal{M}^d$ (Table 1). The polar vector $v^a_{mn}$ satisfies $\mathcal{M}^d v^a_{mn}(\boldsymbol{k}) = (-1)^{\delta_{da}} v^a_{mn}(\boldsymbol{k}')$, where $\boldsymbol{k}'$ is the image of $\boldsymbol{k}$ under $\mathcal{M}^d$ (only the $d$-th component flip its sign), whereas the axial vector $s^i_{mn}$ should satisfy $\mathcal{M}^d s^i_{mn}(\boldsymbol{k}) = -(-1)^{\delta_{di}} s^i_{mn}(\boldsymbol{k}')$. Therefore, one has $\mathcal{M}^d N^{0abc}_{mnl}(\boldsymbol{k}) = -N^{0abc}_{mnl}(\boldsymbol{k}')$ when there are odd number of $d$ within $a, b$ and $c$, and the charge current should vanish in this case. For example, when the system has $\mathcal{M}^x$, then $\sigma^{x,s^0}_{xx}$ and $\sigma^{x,s^0}_{yy}$ should vanish. On the other hand, if $d \neq i$, the spin-$i$ current should vanish when there are even number of $d$ within $a, b$ and $c$, because the $\mathcal{M}^d$ operation on $s^i$ contributes to an additional sign change if $d \neq i$. Therefore, $\sigma^{x,s^z}_{xx}$ and $\sigma^{x,s^z}_{yy}$ could exist in the presence of $\mathcal{M}^x$. Due to the opposite behavior of $N^{0abc}_{mnl}$ and $N^{iabc}_{mnl}$ under $\mathcal{M}^d$, a pure spin current can be generated.

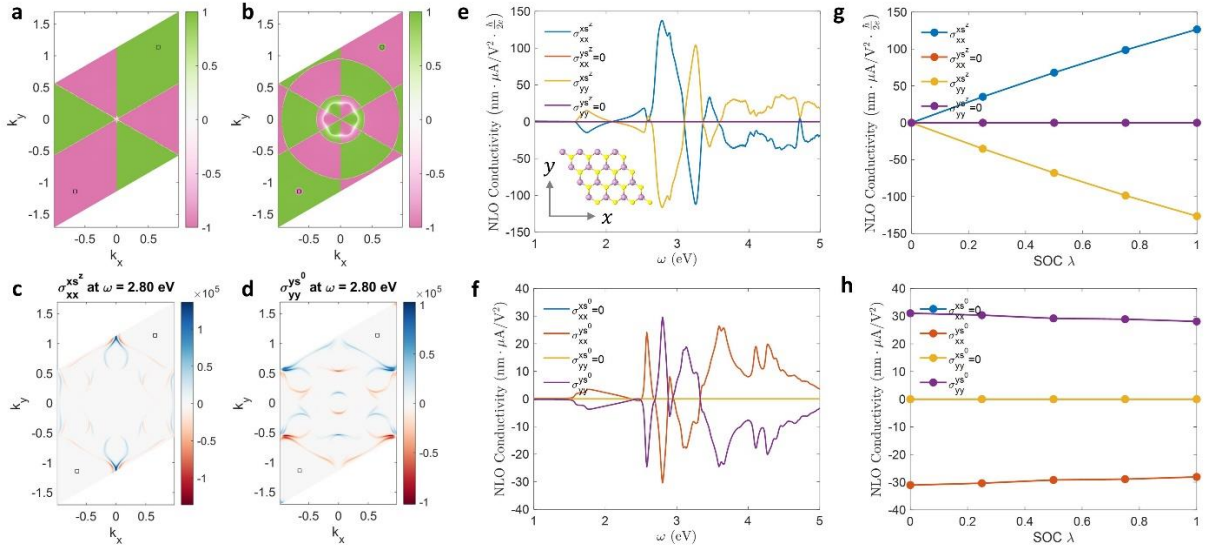

**Figure 2 NLO spin current of MoS$_2$ (a-b)** The spin-$z$ texture $s^z_{mm}(\mathbf{k})$ for the (a) highest valence band and (b) lowest conduction band of MoS$_2$. Nearly all $k$-points have $s^z_{mm}(\mathbf{k}) \cong \pm 1$. **(c-d)** $\mathbf{k}$-specified contribution to the total photoconductivity $\sigma^{xs^z}_{xx}$ and $\sigma^{xs^0}_{xx}$. The black boxes in (a-d) indicate K and K' points in the BZ. **(e-f)** The NLO spin-$z$ and charge conductivity. The complementary behavior is clearly observable: the spin and charge currents are in perpendicular directions. Inset of (e): the atomic structure of MoS$_2$. **(g-h)** Peak values of NLO spin (g) and charge (h) conductivity of MoS$_2$ as a function of SOC strength $\lambda$. The spin conductivity grows linearly with SOC strength, while the charge conductivity is almost independent of SOC strength.

The calculated NLO spin and charge conductivity of monolayer MoS$_2$ under different light polarizations are shown in Figure 2(e-f). One can see that with in-plane polarized light, the nonzero conductivities are complementary for spin and charge currents, consistent with the analysis above. In detail, under the $x$-polarized light, the charge current is along $y$-direction ($\sigma^{x,s^0}_{xx} = 0$ and $\sigma^{y,s^0}_{xx} \neq 0$), whereas the spin-$z$ current is along the $x$-direction ($\sigma^{x,s^z}_{xx} \neq 0$ and $\sigma^{y,s^z}_{xx} = 0$). This indicates that along $x$-direction, equal numbers of spin-up and spin-down electrons are moving oppositely, so the net charge flux is zero, while the net spin flux is nonzero. Along $y$ direction, the spin up and spin down carriers move in the same direction, leading to zero spin current but nonzero charge current (Figure 1). Similar effects occur as well in the case of $y$-polarized light. Interestingly, the spin-$z$ conductivity can be larger than the charge conductivity (in the sense of equivalating $\frac{\hbar}{2} = |e|$). This should be compared with the linear spin Hall effect, where the spin Hall angle (the ratio between the spin conductivity to charge conductivity) is usually on the order of 0.1 and below[49]. We also plot the $\mathbf{k}$-specific contribution to the total conductivity, defined as

$$I_{bc}^{a,s^i}(\omega, \boldsymbol{k}) = \text{Re}\left\{\sum_{mnl} \frac{f_{lm} v_{lm}^b}{E_{ml} - \hbar\omega + i\delta} \left(\frac{j_{mn}^{a,s^i} v_{nl}^c}{E_{mn} + i\delta} - \frac{v_{mn}^c j_{nl}^{a,s^i}}{E_{nl} + i\delta}\right)\right\},$$ in Figure 2(c-d) for $\sigma_{xx}^{x,s^z}$ and $\sigma_{yy}^{y,s^0}$ at $\omega = 2.8$ eV. The mirror symmetry $k_x \to -k_x$ can be clearly observed.

As discussed before, the generation of spin current requires a spin texture. For MoS$_2$, the spin texture is generated by SOC. When SOC is turned off, the spins of electrons are unpolarized, and the spin current would be zero. This is verified by our ab initio calculations. We artificially rescale the strength of SOC in MoS$_2$ by a factor of $\lambda$, and $\lambda = 0$ ($\lambda = 1$) corresponds to no (full) SOC. The NLO conductivities as a function of $\lambda$ are shown in Figure 2(g, h). One can see that when $\lambda = 0$, the spin conductivity is indeed zero. As $\lambda$ increases, the spins would have more and more specified polarization, and the spin conductivity increase accordingly. In contrast, the charge conductivity is nearly independent of $\lambda$.

**Bilayer Anti-ferromagnetic MnBi$_2$Te$_4$.** Next, we study the bilayer AFM MnBi$_2$Te$_4$ (MBT)[50,51], where a large NLO charge current has been reported[52,53]. Each layer of MBT is a septuple layer (SL) in the sequence of Te-Bi-Te-Mn-Te-Bi-Te. The Mn atoms possess magnetic moments $\sim 5~\mu_\text{B}$, with intra-plane ferromagnetic ordering. Bulk MBT is composed of van der Waals (vdW) stacked SLs with inter-plane AFM ordering, and the AFM nature persists when MBT is thinned down to multiple atomic layers. In particular, bilayer MBT is a compensated AFM insulator, whose atomic structure is shown in Figure 3(a). The ground state magnetic moments are pointing along the $z$ direction with a magnetic point group of $\bar{3}'m'$. The atomic structure of bilayer MBT is invariant under $\mathcal{P}$ and the inversion center lies in the vdW gap between the two layers (black square in Figure 3a). However, when one considers magnetism, both $\mathcal{P}$ and $\mathcal{T}$ are broken. Nevertheless, AFM bilayer MBT is invariant under the combined operation $\mathcal{PT}$. Similarly, we find that $\mathcal{PM}^x$ is also preserved. According to the previous analysis (Table 1), we know that $\mathcal{PM}^d v_{mn}^a(\boldsymbol{k}) = -(-1)^{\delta_{ab}} v_{mn}^a(-\boldsymbol{k}')$ and $\mathcal{PM}^d s_{mn}^i(\boldsymbol{k}) = -(-1)^{\delta_{di}} s_{mn}^i(-\boldsymbol{k}')$. Then, one can see that when $d \neq i$, $N^{0abc}$ ($N^{iabc}$) should vanish after Brillouin zone integration when there are even (odd) number of $d$ within $a, b$ and $c$. Therefore, one can still obtain a pure spin current in systems with $\mathcal{PM}^d$ due to the different selection rule on charge and spin currents.

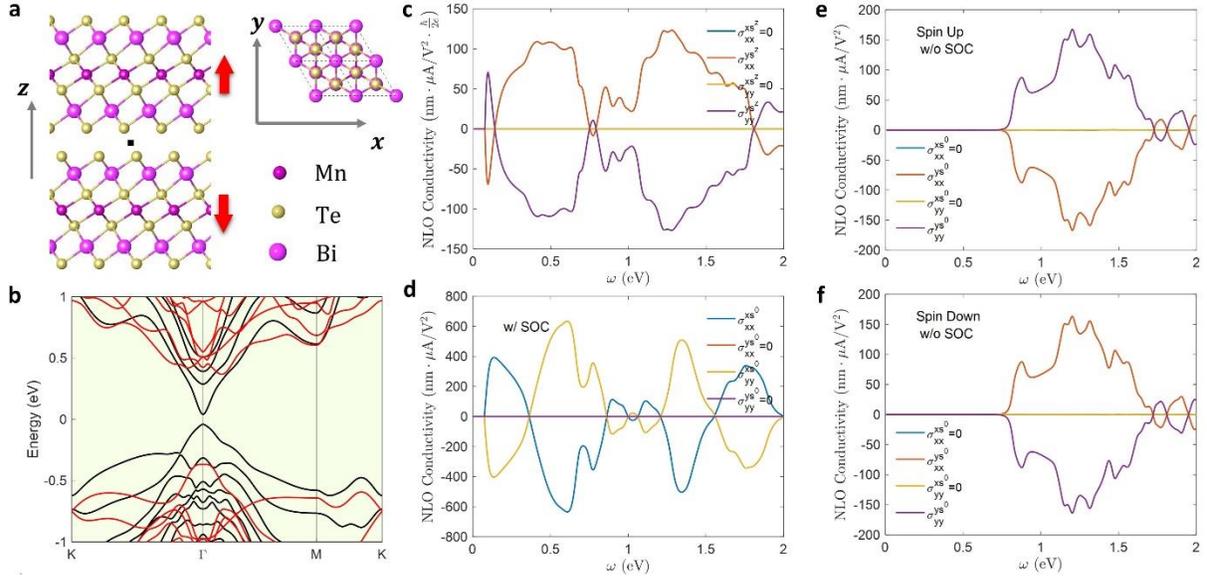

**Figure 3. NLO spin current of MBT (a)** Atomic structure of bilayer MnBi$_2$Te$_4$. The atomic structure has both inversion symmetry $\mathcal{P}$ and mirror symmetry $\mathcal{M}^x$. The inversion center is in between the two layers (black square). The magnetic momentum on Mn is indicated by the red arrows. Considering magnetism, both $\mathcal{P}$ and $\mathcal{M}^x$ break. **(b)** Band structure of MBT with (black) and without (red) SOC. **(c-d)** The NLO spin and charge photoconductivity of bilayer MnBi$_2$Te$_4$ with SOC. Both spin and charge currents have nonzero components and exhibit complementary behavior. **(e-f)** The NLO charge conductivity without SOC. The spin up and down states are treated separately. The photoconductivity from spin-up and down states are exactly opposite to each other. Therefore, the total charge conductivity is zero. But the spin-$z$ conductivity is nonzero.

The band structures of bilayer MBT with and without SOC are shown in Figure 3(b). The bandgap is about 0.1 eV and is located at the Γ point when the SOC effect is included, whereas it is about 0.7 eV and is indirect without SOC. As shown in Figure 3(c-f), the SOC also makes a significant difference in the NLO spin and charge conductivity. When SOC is turned off, $s^z$ is a good quantum number. States with $s^z = \pm 1$ are strictly degenerate in an AFM system and can be treated separately. The NLO conductivities without SOC are shown in Figure 3(e-f), where one can see that the charge current from spin up ($j_\uparrow$) and down ($j_\downarrow$) states are exactly opposite to each other. Consequently, the total charge current $j^{s^0} = j_\uparrow + j_\downarrow$ is zero. However, the total spin-$z$ current $j^{s^z} = j_\uparrow - j_\downarrow$ is nonzero. Therefore, a pure spin current without any charge current is predicted, which comes from the inversion-spin rotation symmetry $\mathcal{PS}$. These results are consistent with those in Ref. [19], where several other well-known AFM materials such as NiO and BiFeO$_3$ were studied.

However, SOC would break $\mathcal{PS}$, and thus lead to a nonzero charge current. Due to the $\mathcal{PM}^x$ symmetry, the charge current is perpendicular to the spin-$z$ current (Figure 3c, d). We also artificially rescale the strength of SOC by a factor of $\lambda$, as done in the MoS$_2$ section (see SM). It is found that the charge conductivity increases with $\lambda$. This is because with a larger $\lambda$, the spin and orbital degrees of freedom are coupled more strongly, and inversion-spin rotation symmetry $\mathcal{PS}$ is broken to a greater extent, thus the charge conductivity would be larger. These results suggest that while SOC enables spin current in non-magnetic materials such as MoS$_2$, it would adversely hinder the generation of pure spin current in some cases. Also, SOC should be treated rigorously when studying both the spin current and the charge current.

**2D Surface of 3D Topological Materials.** Topological insulators[54–56] (TIs) and topological semimetals[57,58] have attracted intense interest in recent years. In TIs, the bulk states are insulating with a finite bandgap, while the surface states are (semi-)metallic with symmetry-protected vanishing bandgap, which has potential applications in electronic and spintronic devices. One salient feature of the surface states is the spin-momentum locking, which could prevent the electrons from backscattering and facilitate spin manipulations[59–61]. In addition, the inversion symmetry $\mathcal{P}$ is naturally broken on the surfaces, even if the bulk possesses $\mathcal{P}$. Therefore, the NLO charge[62] and spin current can be induced solely on surfaces, while the bulk remains silent.

Here we take the topological crystalline insulator (TCI)[63,64] cubic SnTe as an example. The bulk SnTe has space group $Fm\bar{3}m$, and is inversion symmetric inside the 3D crystal, which forbids BPV/BSPV in the bulk interior. But the 2D surfaces of this 3D crystal would lose the inversion symmetry, and therefore can support both BPV and BSPV. Here we consider the {001} surface, which has a four-fold rotational symmetry and double mirror symmetries $\mathcal{M}^x$ and $\mathcal{M}^y$ (Figure 4a). The spectrum function $A(\boldsymbol{k},\omega)$ of the (001) surface is obtained with iterative Green's function method[65,66] and is shown in Figure 4(b-c). In Figure 4(b), $A(\boldsymbol{k},\omega)$ along high symmetry lines in the BZ is presented, and the gapless surface states can be clearly observed. In Figure 4(c), $A(\boldsymbol{k},\omega)$ near $\bar{\text{X}}$ point in the BZ with selected energy $\omega = -0.2, 0,$ and $0.2$ eV are plotted. One can see that $A(\boldsymbol{k},\omega)$ can have significant values on the same $\boldsymbol{k}$-point with different $\omega$, enabling strong interband transitions and significant photocurrents. In addition, the surface spin textures are plotted as black arrows. The nonzero $s^x$ and $s^y$ components indicate that one can obtain spin-$x$ and spin-$y$ currents.

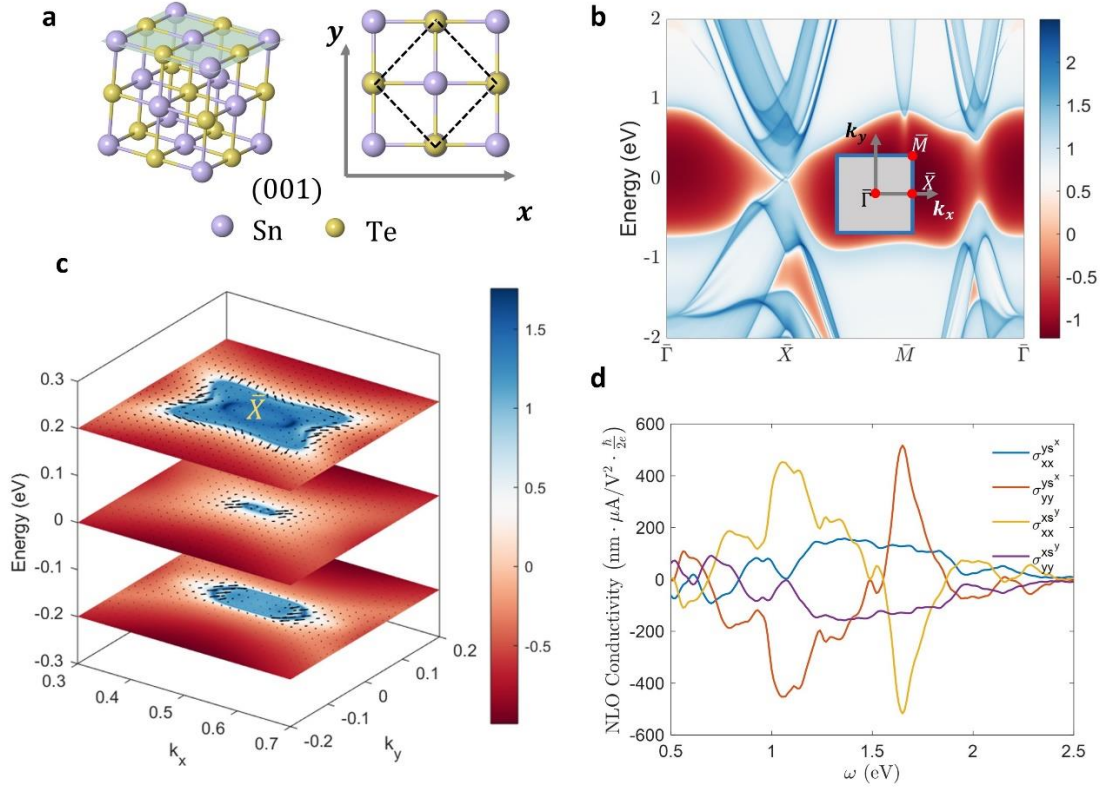

**Figure 4. NLO spin current on the (001) surface of SnTe. (a)** The atomic structure of SnTe. In the left panel the (001) face is painted in light green, which possesses double mirror symmetries $\mathcal{M}^x$ and $\mathcal{M}^y$. The dashed box in the right panel indicates the primitive cell on the surface. **(b)** The surface spectrum function $A(\mathbf{k}, \omega)$ on high symmetry lines in the BZ. The gapless surface states can be observed. **(c)** The surface spectrum function $A(\mathbf{k}, \omega)$ over the BZ for selected $\omega = -0.2, 0$, and $0.2\ eV$. $k_x$ and $k_y$ are in the unit of reciprocal lattices. The surface spin textures are indicated by the black arrows. Color scheme (red to blue) in (b) and (c) represents surface state contribution. The color bars are in logarithmic scale, and the energy is offset to the valence band maximum. **(d)** The NLO spin current conductivity with $x$ and $y$ spin polarizations. Note that all charge and spin-$z$ current components are vanishing due to $\mathcal{M}^x$ and $\mathcal{M}^y$.

According to our previous symmetry analysis, under in-plane polarized light ($b, c = x$ or $y$), no NLO charge or spin-$z$ current can be generated on the $\{001\}$ surface, due to the double mirror symmetry $\mathcal{M}_x$ and $\mathcal{M}_y$. However, it is possible to have nonzero spin-$x$ and spin-$y$ currents, which is verified by our ab initio calculations. We use a slab model to compute the surface NLO spin and charge conductivity. To distinguish the contribution from only one surface of the slab, we define a projection operator[67] $P_l = \sum_{i \in l} |\psi_i\rangle\langle\psi_i|$. Here $|\psi_i\rangle$ are atomic orbitals located on the $l$-th atomic layer. Then we replace the current operator $j$ in Equation (2) with $P_l j P_l$, and the resultant conductivity can be layer distinguished (on the $l$-th layer). Note that there could be nonzero cross-terms $P_l j P_m$ (with $l \neq m$), indicating the interference between the $l$-th and $m$-th layer. From our computations, even for neighboring layers with $m = l \pm 1$, the contribution from $P_l j P_m$ is quite small (less than 10 %). Here for a conceptual demonstration of our theory, we only consider $P_{l=1} j P_{l=1}$ and calculate the contribution from the out-most layer. NLO spin-$x$ and spin-$y$ conductivities are plotted in Figure 4(d). One can see that the maximum value of $\sigma_{yy}^{ys^x}$ can reach 500 nm · µV/Å² · $\frac{\hbar}{2e}$. We would like to emphasize again that under the light field with in-plane polarization, NLO charge current is absent on this {001} surface, therefore a pure spin current without any charge current can be generated due to the double mirror symmetries. Such methodology can also be used to distinguish surface and bulk states and to probe the surface states. There may be other systems that possess double mirror symmetries as well, such as monolayer FeSe[68], that may be good candidates for pure spin current generation.

**Table 2** Mechanisms for NLO charge and spin photocurrent generation under different material symmetries and light polarizations. For the shift mechanism, the conductivity contribution is independent of the carrier lifetime $\tau$. Whereas for the injection mechanism, the conductivity contribution scales linearly with $\tau$.

|  | $\mathcal{P}$-conserved | $\mathcal{P}$-broken $\mathcal{T}$-conserved | $\mathcal{P}$-broken, $\mathcal{T}$-broken $\mathcal{PT}$-conserved | $\mathcal{P}$-broken, $\mathcal{T}$-broken $\mathcal{PT}$-broken |
|---|---|---|---|---|
| DC Charge Current (BPV) | No | LPL ⇔ Shift<br>CPL ⇔ Injection | LPL ⇔ Injection<br>CPL ⇔ Shift+injection | LPL ⇔ Shift+Injection<br>CPL ⇔ Shift+Injection |
| DC Spin Current (BSPV) | No | LPL ⇔ Injection<br>CPL ⇔ Shift+injection | LPL ⇔ Shift<br>CPL ⇔ Injection | LPL ⇔ Shift+Injection<br>CPL ⇔ Shift+Injection |

## Discussions

Before concluding, we would like to make some remarks. First, it is well known that BPV has potentially shift and injection current contributions. The shift mechanism comes from the fact that the wavefunction center of the electron and hole band states are different, leading to an electric dipole upon photon absorption. On the other hand, the injection mechanism comes from the fact that the electron and holes have different velocities, and that the coherent $\boldsymbol{k}$ and $-\boldsymbol{k}$ excitations are imbalanced, leading to $\boldsymbol{k}$ and $-\boldsymbol{k}$ asymmetry in steady-state population and a net current. These facts are more evident if we transform Equation (2) into the length gauge, as shown in the SM. In a $\mathcal{T}$-conserved system, the DC charge currents under LPL and CPL have shift and injection mechanism, respectively[35]. In contrast, for the DC spin current, the mechanism under LPL and CPL should be injection-like and (shift+injection)-like (see SM). Here the shift- (injection-) current is defined by the conductivity scaling relationship as $\propto \tau^0$ ($\tau^1$). Therefore, the spin conductivity in Figure 2e and Figure 4d can be further enhanced if a larger $\tau$ is used (see SM). The different mechanisms for spin and charge current come from the different behavior of $N_{mnl}^{iabc}$ ($i \neq 0$) and $N_{mnl}^{0abc}$ under $\mathcal{T}$-operation. Note that in $\mathcal{T}$-conserved systems, the shift spin current should vanish under LPL, consistent with the arguments in Ref. [20]. We have done similar analyses on mechanisms of current generation under different symmetry conditions, and the results are listed in Table 2. These results are also computationally verified by varying $\tau$ (see details in SM).

Second, as shown above, a pure spin current induced by mirror symmetry is usually accompanied by a charge current in the transverse direction (except for the (100) surface states of cubic SnSe, with double mirror symmetry $\mathcal{M}^x$ and $\mathcal{M}^y$). It is possible the achieve a pure spin current without any charge current at all, if the system possesses inversion-spin rotation symmetry $\mathcal{PS}$. One can see that $\mathcal{PS} N_{mnl}^{0abc}(\mathbf{k}) = -\widehat{N}_{mnl}^{0abc}(-\mathbf{k})$, where $\widehat{\cdot}$ indicates $\cdot$ obtained on the spin partner state. Therefore, the charge current should identically be zero in the presence of $\mathcal{PS}$. On the other hand, $\mathcal{PS} N_{mnl}^{iabc}(\mathbf{k}) = -e^{i\phi}\widehat{N}_{mnl}(-\mathbf{k})$, where $e^{i\phi}$ is a phase factor induced by the spin rotation operation on $s^i$. Thus, the spin current does not have to vanish. In fact, $\mathcal{PS}_z$, where $\mathcal{S}_z$ flips the spin up and down states, is the origin of the vanishing charge current of MBT when SOC is ignored. In practice, a skyrmion lattice, or magnetic materials with canted or all-in-all-out magnetic ordering can be an ideal platform for the generation of pure spin current without any charge current.

Third, the NLO conductivity in Equation (2) is obtained from the quadratic response theory. It essentially is $\text{Tr}(j^{(0)}\rho^{(2)})$, where $j^{(0)}$ is the current operator independent of the electric field $E$, while $\rho^{(2)}$ is the second-order perturbation in the density matrix and is proportional to $E^2$. There could be other mechanisms for the generation of spin/charge current. For example, there could be an anomalous velocity, which leads to an additional term $j^{(1)}$ in the current operator that is linearly dependent on $E$. $j^{(1)}$ can contribute to an NLO conductivity from $\text{Tr}(j^{(1)}\rho^{(1)})$, where $\rho^{(1)}$ is the first-order perturbation in the density matrix. Note that this contribution should vanish for all the material systems studied in this work.

Finally, we would like to briefly discuss how the spin current can be detected. There are well-established approaches for detecting the spin current generated by, e.g., spin Hall effect[9]. For example, with an open circuit setup (SM Figure S2a), the spin would accumulate on the ends of the source material. The spin accumulation can be measured by magneto-optic effects such as Kerr rotation or Faraday effect[69]. Also, in a close circuit setup (SM Figure S2b), the spin current source is sandwiched between two metallic leads (e.g., Pt). The light-induced spin current is transmitted to the metallic leads. An inverse spin Hall voltage would be generated transverse to the spin current[70–72], and the spin current can be measured by the inverse spin Hall voltage. Assuming a spin conductivity of 100 µA/V$^2$ $\frac{\hbar}{2e}$, an electric field as small as 100 V/m would generate a spin current density of $1 \frac{A}{m^2} \frac{\hbar}{2e}$. Assume a spin Hall angle of 0.1, the current density in the metallic lead would be 10 A/m$^2$, which can be detectable.

In conclusion, we demonstrate a generic picture of spin photocurrent generation with nonlinear light-matter interactions. By symmetry analysis, we reveal that the effect does not have any special requirements, except for the inversion symmetry breaking. Thus, it applies to a wide range of materials and extended defects like surfaces, stacking faults, grain boundaries, and dislocations. If the system possesses mirror symmetry or inversion-mirror symmetry, a pure spin current can be realized. Our theory is verified with ab initio calculations in several material systems, and the spin current conductivity is found to be comparable or even bigger than its charge BPV cousin. The predicted BSPV platforms can be readily integrated with existing semiconductor technologies. They may find applications in next-generation ultrafast spintronics and quantum information processing.

## Methods

The first-principles calculations are based on density functional theory (DFT)[73,74] as implemented in Vienna *ab initio* simulation package (VASP)[75,76]. The exchange-correlation interactions are treated by a generalized gradient approximation (GGA) in the form of Perdew-Burke-Ernzerhof (PBE)[77]. Core and valence electrons are treated by projector augmented wave (PAW) method[78] and plane-wave basis functions, respectively. For DFT calculations, the first Brillouin zone is sampled by a Γ-centered ***k***-mesh with grid density of at least $2\pi \times 0.02$ Å$^{-1}$ along each dimension. The DFT+U method is adopted to treat the $d$ orbitals of Mn atoms in MBT ($U = 4.0$ eV). Then a tight-binding (TB) Hamiltonian is constructed from DFT results with the help of the Wannier90 package[79]. The TB Hamiltonian is utilized to calculate the NLO charge and spin conductivity according to Equation (2) on a finer $\mathbf{k}$-mesh. The $\mathbf{k}$-mesh convergence for BZ integration is well tested. In practice, the BZ integration in Equation (2) is carried out by $\mathbf{k}$-mesh sampling with $\int \frac{dk}{(2\pi)^3} = \frac{1}{V}\sum_k w_k$, where $V$ is the volume of the simulation cell in real space and $w_k$ is weight factor. However, for 2D materials, the definition of volume $V$ is ambiguous, because the thickness of 2D materials is ambiguously defined[80]. Thus, we replace volume $V$ with the area $S$, and the 2D and 3D conductivity satisfy $\sigma_{2D} = L\sigma_{3D}$, where $L$ is an effective thickness of the material.

**Acknowledgments.** This work was supported by an Office of Naval Research MURI through grant #N00014-17-1-2661. We are grateful for the insightful suggestions by Dr. Zhurun Ji.

**Data availability.** The authors declare that the main data supporting the findings of this study are available within the article and its Supplementary Information files.

**Code availability.** The MATLAB code for computing the NLO conductivities is available from the corresponding author upon reasonable request.

**Author contributions.** H.X. and J.L. conceived the idea and designed the project. H.X. derived the theories. H.X. performed the calculations and wrote the paper with the help of H.W. and J.Z. J.L supervised the project. All authors analyzed the data and contributed to the discussions of the results.

**Competing Interests.** The authors declare no competing interests.

## References

1. Žutić, I., Fabian, J. & Sarma, S. Das. Spintronics: Fundamentals and applications. *Reviews of Modern Physics* **76**, 323–410 (2004).
2. Bader, S. D. & Parkin, S. S. P. Spintronics. *Annu. Rev. Condens. Matter Phys.* **1**, 71–88 (2010).
3. Shor, P. W. Algorithms for quantum computation: discrete logarithms and factoring. in 124–134 (Institute of Electrical and Electronics Engineers (IEEE), 2002). doi:10.1109/sfcs.1994.365700
4. Ladd, T. D. *et al.* Quantum computers. *Nature* **464**, 45–53 (2010).
5. Umesh, S. & Mittal, S. A survey of spintronic architectures for processing-in-memory and neural networks.

*Journal of Systems Architecture* **97**, 349–372 (2019).

6. Awschalom, D. D. & Flatté, M. E. Challenges for semiconductor spintronics. *Nature Physics* **3**, 153–159 (2007).
7. Ganichev, S. D. *et al.* Spin-galvanic effect. *Nature* **417**, 153–156 (2002).
8. Benítez, L. A. *et al.* Tunable room-temperature spin galvanic and spin Hall effects in van der Waals heterostructures. *Nat. Mater.* **19**, 170–175 (2020).
9. Sinova, J., Valenzuela, S. O., Wunderlich, J., Back, C. H. & Jungwirth, T. Spin Hall effects. *Rev. Mod. Phys.* **87**, 1213–1260 (2015).
10. D'Yakonov, M. I., Perel', V. I., D'Yakonov, M. I. & Perel', V. I. Possibility of Orienting Electron Spins with Current. *JETPL* **13**, 467 (1971).
11. Dyakonov, M. I. & Perel, V. I. Current-induced spin orientation of electrons in semiconductors. *Phys. Lett. A* **35**, 459–460 (1971).
12. Ando, K. *et al.* Inverse spin-Hall effect induced by spin pumping in metallic system. in *Journal of Applied Physics* **109**, 103913 (American Institute of PhysicsAIP, 2011).
13. Uchida, K. *et al.* Observation of the spin Seebeck effect. *Nature* **455**, 778–781 (2008).
14. Adachi, H., Uchida, K., Saitoh, E. & Maekawa, S. Theory of the spin Seebeck effect. *Reports Prog. Phys.* **76**, 036501 (2013).
15. Sheng, P. *et al.* The spin Nernst effect in tungsten. *Sci. Adv.* **3**, e1701503 (2017).
16. Meyer, S. *et al.* Observation of the spin Nernst effect. *Nat. Mater.* **16**, 97–981 (2017).
17. Ganichev, S. D. *et al.* Conversion of spin into directed electric current in quantum wells. *Phys. Rev. Lett.* **86**, 4358–4361 (2001).
18. Luo, Y. K. *et al.* Opto-valleytronic spin injection in monolayer MoS2/few-layer graphene hybrid spin valves. *Nano Lett.* **17**, 3877–3883 (2017).
19. Young, S. M., Zheng, F. & Rappe, A. M. Prediction of a linear spin bulk photovoltaic effect in antiferromagnets. *Phys. Rev. Lett.* **110**, 057201 (2013).
20. Kim, K. W., Morimoto, T. & Nagaosa, N. Shift charge and spin photocurrents in Dirac surface states of topological insulator. *Phys. Rev. B* **95**, 035134 (2017).
21. Ishizuka, H. & Sato, M. Rectification of Spin Current in Inversion-Asymmetric Magnets with Linearly Polarized Electromagnetic Waves. *Phys. Rev. Lett.* **122**, 197702 (2019).
22. Ishizuka, H. & Sato, M. Theory for shift current of bosons: Photogalvanic spin current in ferrimagnetic and antiferromagnetic insulators. *Phys. Rev. B* **100**, 224411 (2019).
23. Žutić, I., Fabian, J. & Das Sarma, S. Proposal for a spin-polarized solar battery. *Appl. Phys. Lett.* **79**, 1558–1560 (2001).
24. Žutić, I., Fabian, J., Fabian, J. & Das Sarma, S. Spin-polarized transport in inhomogeneous magnetic semiconductors: Theory of magnetic/nonmagnetic p-n junctions. *Phys. Rev. Lett.* **88**, 66603/1-66603/4 (2002).
25. Endres, B. *et al.* Demonstration of the spin solar cell and spin photodiode effect. *Nat. Commun.* **4**, 1–5

(2013).

26. Stevens, M. J. *et al.* Quantum Interference Control of Ballistic Pure Spin Currents in Semiconductors. *Phys. Rev. Lett.* **90**, 4 (2003).

27. Hübner, J. *et al.* Direct Observation of Optically Injected Spin-Polarized Currents in Semiconductors. *Phys. Rev. Lett.* **90**, 4 (2003).

28. Hamamoto, K., Ezawa, M., Kim, K. W., Morimoto, T. & Nagaosa, N. Nonlinear spin current generation in noncentrosymmetric spin-orbit coupled systems. *Phys. Rev. B* **95**, 224430 (2017).

29. Belinicher, V. I. & Sturman, B. I. The photogalvanic effect in media lacking a center of symmetry. *Sov. Phys. - Uspekhi* **23**, 199–223 (1980).

30. Von Baltz, R. & Kraut, W. Theory of the bulk photovoltaic effect in pure crystals. *Phys. Rev. B* **23**, 5590–5596 (1981).

31. Fert, A. Nobel lecture: Origin, development, and future of spintronics nobel lecture: Origin, development, and future of spintronics. *Rev. Mod. Phys.* **80**, 1517–1530 (2008).

32. Puebla, J., Kim, J., Kondou, K. & Otani, Y. Spintronic devices for energy-efficient data storage and energy harvesting. *Commun. Mater.* **1**, 1–9 (2020).

33. Bernevig, B. A. & Zhang, S. C. Quantum spin hall effect. *Phys. Rev. Lett.* **96**, 106802 (2006).

34. Kraut, W. & Von Baltz, R. Anomalous bulk photovoltaic effect in ferroelectrics: A quadratic response theory. *Phys. Rev. B* **19**, 1548–1554 (1979).

35. Sipe, J. & Shkrebtii, A. Second-order optical response in semiconductors. *Phys. Rev. B - Condens. Matter Mater. Phys.* **61**, 5337–5352 (2000).

36. Ventura, G. B., Passos, D. J., Lopes Dos Santos, J. M. B., Viana Parente Lopes, J. M. & Peres, N. M. R. Gauge covariances and nonlinear optical responses. *Phys. Rev. B* **96**, 035431 (2017).

37. Taghizadeh, A., Hipolito, F. & Pedersen, T. G. Linear and nonlinear optical response of crystals using length and velocity gauges: Effect of basis truncation. *Phys. Rev. B* **96**, 195413 (2017).

38. Sinova, J. *et al.* Universal intrinsic spin Hall effect. *Phys. Rev. Lett.* **92**, 126603 (2004).

39. Rashba, E. I. Spin currents in thermodynamic equilibrium: The challenge of discerning transport currents. *Phys. Rev. B - Condens. Matter Mater. Phys.* **68**, 241315 (2003).

40. Shi, J., Zhang, P., Xiao, D. & Niu, Q. Proper definition of spin current in spin-orbit coupled systems. *Phys. Rev. Lett.* **96**, 076604 (2006).

41. Sun, Q. F., Xie, X. C. & Wang, J. Persistent spin current in nanodevices and definition of the spin current. *Phys. Rev. B - Condens. Matter Mater. Phys.* **77**, 035327 (2008).

42. Fei, R., Lu, X. & Yang, L. Intrinsic Spin Photogalvanic Effect in Nonmagnetic Insulator. (2020).

43. Smit, J. The spontaneous hall effect in ferromagnetics I. *Physica* **21**, 877–887 (1955).

44. Smit, J. The spontaneous hall effect in ferromagnetics II. *Physica* **24**, 39–51 (1958).

45. Berger, L. Influence of spin-orbit interaction on the transport processes in ferromagnetic nickel alloys, in the presence of a degeneracy of the 3d band. *Physica* **30**, 1141–1159 (1964).

46. Berger, L. Side-jump mechanism for the hall effect of ferromagnets. *Phys. Rev. B* **2**, 4559–4566 (1970).

47. Zeng, H. *et al.* Optical signature of symmetry variations and spin-valley coupling in atomically thin tungsten dichalcogenides. *Sci. Rep.* **3**, 1–5 (2013).

48. Bawden, L. *et al.* Spin-valley locking in the normal state of a transition-metal dichalcogenide superconductor. *Nat. Commun.* **7**, 1–6 (2016).

49. Gradhand, M., Fedorov, D. V., Zahn, P. & Mertig, I. Spin Hall angle versus spin diffusion length: Tailored by impurities. *Phys. Rev. B - Condens. Matter Mater. Phys.* **81**, 245109 (2010).

50. Li, J. *et al.* Intrinsic magnetic topological insulators in van der Waals layered MnBi2Te4-family materials. *Sci. Adv.* **5**, eaaw5685 (2019).

51. Deng, Y. *et al.* Quantum anomalous Hall effect in intrinsic magnetic topological insulator MnBi2Te4. *Science (80-. ).* **367**, 895–900 (2020).

52. Wang, H. & Qian, X. Giant nonlinear photocurrent in $\mathcal{PT}$-symmetric magnetic topological quantum materials. (2020).

53. Fei, R., Song, W. & Yang, L. Giant linearly-polarized photogalvanic effect and second harmonic generation in two-dimensional axion insulators. (2020).

54. Hasan, M. Z. & Kane, C. L. Colloquium: Topological insulators. *Rev. Mod. Phys.* **82**, 3045–3067 (2010).

55. Qi, X. L. & Zhang, S. C. Topological insulators and superconductors. *Rev. Mod. Phys.* **83**, 1057 (2011).

56. Bansil, A., Lin, H. & Das, T. Colloquium: Topological band theory. *Rev. Mod. Phys.* **88**, 021004 (2016).

57. Armitage, N. P., Mele, E. J. & Vishwanath, A. Weyl and Dirac semimetals in three-dimensional solids. *Rev. Mod. Phys.* **90**, 015001 (2018).

58. Yan, B. & Felser, C. Topological Materials: Weyl Semimetals. *Annu. Rev. Condens. Matter Phys.* **8**, 337–354 (2017).

59. Tian, J., Hong, S., Miotkowski, I., Datta, S. & Chen, Y. P. Observation of current-induced, long-lived persistent spin polarization in a topological insulator: A rechargeable spin battery. *Sci. Adv.* **3**, e1602531 (2017).

60. Mellnik, A. R. *et al.* Spin-transfer torque generated by a topological insulator. *Nature* **511**, 449–451 (2014).

61. Lin, B. C. *et al.* Electric Control of Fermi Arc Spin Transport in Individual Topological Semimetal Nanowires. *Phys. Rev. Lett.* **124**, 116802 (2020).

62. Chang, G. *et al.* Unconventional Photocurrents from Surface Fermi Arcs in Topological Chiral Semimetals. *Phys. Rev. Lett.* **124**, 166404 (2020).

63. Fu, L. Topological crystalline insulators. *Phys. Rev. Lett.* **106**, 106802 (2011).

64. Hsieh, T. H. *et al.* Topological crystalline insulators in the SnTe material class. *Nat. Commun.* **3**, 1–7 (2012).

65. Sancho, M. P. L., Sancho, J. M. L. & Rubio, J. Quick iterative scheme for the calculation of transfer matrices: application to Mo (100). *J. Phys. F Met. Phys.* **14**, 1205 (1984).

66. Sancho, M. P. L., Sancho, J. M. L., Sancho, J. M. L. & Rubio, J. Highly convergent schemes for the calculation of bulk and surface Green functions. *J. Phys. F Met. Phys.* **15**, 851 (1985).

67. Varnava, N. & Vanderbilt, D. Surfaces of axion insulators. *Phys. Rev. B* **98**, 245117 (2018).

68. Huang, D. & Hoffman, J. E. Monolayer FeSe on $SrTiO_3$. *Annu. Rev. Condens. Matter Phys.* **8**, 311–336 (2017).
69. Kato, Y. K., Myers, R. C., Gossard, A. C. & Awschalom, D. D. Observation of the spin hall effect in semiconductors. *Science (80-. ).* **306**, 1910–1913 (2004).
70. Saitoh, E., Ueda, M., Miyajima, H. & Tatara, G. Conversion of spin current into charge current at room temperature: Inverse spin-Hall effect. *Appl. Phys. Lett.* **88**, 182509 (2006).
71. Valenzuela, S. O. & Tinkham, M. Direct electronic measurement of the spin Hall effect. *Nature* **442**, 176–179 (2006).
72. Kimura, T., Otani, Y., Sato, T., Takahashi, S. & Maekawa, S. Room-temperature reversible spin hall effect. *Phys. Rev. Lett.* **98**, 156601 (2007).
73. Hohenberg, P. & Kohn, W. Inhomogeneous Electron Gas. *Phys. Rev.* **136**, B864–B871 (1964).
74. Kohn, W. & Sham, L. J. Self-Consistent Equations Including Exchange and Correlation Effects. *Phys. Rev.* **140**, A1133–A1138 (1965).
75. Kresse, G. & Furthmüller, J. Efficiency of ab-initio total energy calculations for metals and semiconductors using a plane-wave basis set. *Comput. Mater. Sci.* **6**, 15–50 (1996).
76. Kresse, G. & Furthmüller, J. Efficient iterative schemes for *ab initio* total-energy calculations using a plane-wave basis set. *Phys. Rev. B* **54**, 11169–11186 (1996).
77. Perdew, J. P., Burke, K. & Ernzerhof, M. Generalized Gradient Approximation Made Simple. *Phys. Rev. Lett.* **77**, 3865–3868 (1996).
78. Blöchl, P. E. Projector augmented-wave method. *Phys. Rev. B* **50**, 17953–17979 (1994).
79. Mostofi, A. A. *et al.* An updated version of wannier90: A tool for obtaining maximally-localised Wannier functions. *Comput. Phys. Commun.* **185**, 2309–2310 (2014).
80. Laturia, A., Van de Put, M. L. & Vandenberghe, W. G. Dielectric properties of hexagonal boron nitride and transition metal dichalcogenides: from monolayer to bulk. *npj 2D Mater. Appl.* **2**, 1–7 (2018).